\documentclass[
aip,
reprint,
groupedaddress
]{revtex4-2}

\usepackage{siunitx}
\DeclareSIUnit[number-unit-product = \,]{\atpercent}{at.\%} 

\usepackage{graphicx}
\usepackage[outdir=./]{epstopdf}
\usepackage{amsmath, amssymb}
\usepackage{booktabs}
\usepackage{nicefrac}

\newcommand{\lwr}[1]{\textsubscript{\protect\raisebox{-1pt}{#1}}}

\newcommand{\mlwr}[1]{_{\mathrm{#1}}}			
\newcommand{\mupr}[1]{^{\mathrm{#1}}}

\newcommand{\diff}{\mathrm{d}}
\newcommand{\imag}{\text{i}}

\begin{document}
	
	\title[Identification of acceptors in NiO]{Identification of Li\lwr{Ni} and V\lwr{Ni} acceptor levels in doped nickel oxide}
	\author{Robert Karsthof}
	\affiliation{Universit\"{a}t Leipzig, Felix-Bloch-Institut f\"{u}r Festkörperphysik, Linn\'{e}str. 5, 04103 Leipzig, Germany}
	\email{r.m.karsthof@smn.uio.no}
	\altaffiliation{Centre for Materials Science and Nanotechnology, Universitetet i Oslo, Gaustadalléen 23A, 0373 Oslo, Norway}

	\author{Holger von Wenckstern}
	\affiliation{Universit\"{a}t Leipzig, Felix-Bloch-Institut f\"{u}r Festkörperphysik, Linn\'{e}str. 5, 04103 Leipzig, Germany}
	
	\author{Marius Grundmann}
	\affiliation{Universit\"{a}t Leipzig, Felix-Bloch-Institut f\"{u}r Festkörperphysik, Linn\'{e}str. 5, 04103 Leipzig, Germany}

	\date{\today}
	
\begin{abstract}
	Nickel oxide, in particular in its doped, semiconducting form, is an important component of several opto-electronic devices. Doping NiO is commonly achieved either by incorporation of lithium, which readily occupies Ni sites substitutionally, producing the Li\lwr{Ni} acceptor, or by supplying reactive oxygen species during NiO film deposition, which leads to the formation of Ni vacancies (V\lwr{Ni}). However, the energetic position of these acceptors in the NiO band gap has not been experimentally determined until today. In this work, we close this knowledge gap by studying rectifying $n\mupr{++}p$ heterojunctions of NiO on top of fluorine-doped tin oxide. These structures show sufficient rectification to perform electric characterization by defect spectroscopic techniques, specifically capacitance-voltage and thermal admittance spectroscopy. Using these methods, the $(0/-)$ charge transition levels are determined to be \SI{190}{\milli\electronvolt} and \SI{409}{\milli\electronvolt} above the valence band edge for the Li\lwr{Ni} and the V\lwr{Ni} acceptor, respectively.
\end{abstract}

\maketitle

\section{Introduction}
	
Nickel oxide is one of the rare examples of a $p$-type semiconducting metal oxide. Because of its high band gap of \SI{3.8}{\electronvolt}, it is additionally transparent in the visible spectral range, making it interesting for a variety of optoelectronic bipolar devices, like organic \cite{He.1999,Irwin.2008,Park.2010} and perovskite solar cells \cite{Jeng.2014}, light-emitting diodes \cite{Park.2005,Tang.2013}, electrochromic devices \cite{Avendano.2006,Huang.2011,Moulki.2012,Ren.2013}, and resistive-switching elements \cite{Seo.2005,Kim.2008}. For many of these applications, doping is a key element in obtaining functional layers and devices since pure, stoichiometric NiO is known to be a good insulator. The knowledge of the energetic positions of acceptor states with respect to the band edges is important when it comes to tayloring NiO properties to the respective device. For example, heavy doping may lead to Fermi level pinning, which in turn impacts the band alignment in heterostructures. Acceptor doping of NiO is typically achieved by either incorporating Li into the NiO lattice, which substitutes for Ni and contributes a single hole to the electronic system, or by growing the sample under an oversupply of oxygen which renders the specimen Ni-deficient, producing nickel vacancy (V\lwr{Ni}) double acceptors. Even though there exist some (mostly theoretical) studies that estimate the positions of the charge transition levels associated with these acceptors (see Table~\ref{tab:lit_act_energies}), no experimental verification of these estimates has been published to date. To our view, this is a result of a lack of rectifying NiO-based structures, like Schottky or heterodiodes, with a depletion layer located within the NiO layer that enables the use of conventional defect spectroscopical techniques, like thermal admittance spectroscopy (TAS). This is mainly due to the low electron affinity of NiO of only around \SI{1.5}{\electronvolt} \cite{Wu.1997}, limiting the choice of materials with even lower work functions to achieve hole depletion.

With this work, we show that $n\mupr{++}$-FTO/$p$-NiO heterocontacts possess the desired properties in that they lead to a considerable depletion layer width of 25-\SI{35}{\nano\meter} at \SI{0}{\volt} bias, and a band bending of roughly \SI{0.5}{\volt}, making capacitance-voltage measurements and thermal admittance spectroscopy and the determination of the acceptor charge transition levels possible. At the same time, the current study can be considered a test case for these techniques regarding their ability to reliably probe trap levels in materials with extremely low hole mobilities.

\begin{table}
\caption{Literature values for the expected charge transition levels of the Li cation-substitutional and the Ni vacancy acceptors. All values in \si{\milli\electronvolt} with respect to valence band maximum.}
\label{tab:lit_act_energies}
\begin{tabular}{S[table-format=4.0]S[table-format=4.0]S[table-format=4.0]c} \toprule
	 \multicolumn{1}{c}{Li\lwr{Ni}} & \multicolumn{2}{c}{V\lwr{Ni}}  & method/ ref.  \\ \cmidrule{2-3}
	{$(0/-)$} & {$(0/-)$} & {$(-/2-)$} & \\ \midrule 
 	\num{250}  & \num{480} & \num{720} & theor. \cite{Lany.2007} \\
 	\num{153} & \num{515} & & theor. \cite{Wu.2009} \\
 	 & \num{260} & \num{370} & theor. \cite{Zhang.2008} \\
 	 & \num{-30} & \num{160} & theor. \cite{Lee.2010} \\
 	\num{400} & \num{600} & \num{800} & theor. \cite{Adler.1970} \\
 	 & \num{400} &  \num{600} & theor \cite{Dawson.2015} \\
 	 \midrule
 	 \num{190} & \num{409} & & exp. (this work) \\
	\bottomrule
\end{tabular}
\end{table}

\section{Experimental methods}

NiO films were deposited on top of glass substrates coated with fluorine-doped tin oxide (FTO), purchased from Calyxo GmbH, Germany ($\sigma\mlwr{FTO} = \SI{2.5e3}{\siemens\per\centi\meter}$, $n\mlwr{FTO,Hall} = \SI{5.5e20}{\per\cubic\centi\meter}$). NiO growth was performed by pulsed laser deposition (PLD) using a KrF excimer laser (wavelength \SI{248}{\nano\meter}, pulse energy \SI{650}{\milli\joule}) ablating either ceramic targets of either pure (purity \SI{99.998}{\percent}) or \SI{0.5}{\atpercent} Li-doped NiO (from mixing Li\lwr{2}CO\lwr{3} and NiO powders, Alfa Aesar). The growth was done either at room temperature (no intentional substrate heating) or at \SI{300}{\degreeCelsius}. The partial pressure of the background oxygen gas was chosen between \SI{0.02}{\milli\\bar} for Li-doping and \SI{0.1}{\milli\bar} for introducing an excess of oxygen during growth. Thicknesses of the NiO films were between \SI{100}{\nano\meter} and \SI{290}{\nano\meter}. The NiO layers were capped by a \SI{20}{\nano\meter} thick Pt layer for uniform current flow, deposited by DC magnetron sputtering. Additionally to PLD, NiO growth was also achieved by reactive DC magnetron sputtering from a metallic Ni target at room temperature in an Ar/O\lwr{2} atmosphere ($p\mlwr{O2}= \SI{0.03}{\milli\bar}$), capped in the same way as the PLD NiO layers. As was shown in Ref.~\cite{Karsthof.2019}, Pt/NiO contacts do not induce a depletion region in NiO and can be considered ohmic for small voltages around \SI{0}{\volt}.

For the room temperature-grown NiO, the NiO/Pt stack was patterned into pillars with circular cross sections (diameter between \SI{250}{\micro\meter} and \SI{800}{\micro\meter}) by deposition of the stack though a previously structured photoresist mask. For the high temperature-grown NiO film, only the Pt contacts were patterned in the same way. 

Current-voltage characterization was carried out with an AGILENT 4155C semiconductor parameter analyzer connected to a Süss wafer prober system using tungsten needles. These measurements were used to select individual contacts for the defect spectroscopy measurements. The samples were mounted onto transistor sockets, and the selected pillars were contacted with Au wires and conductive epoxy resin. Defect spectroscopic measurements were carried out in a He flow cryostat ($\SI{20}{\kelvin} < T\mlwr{meas} < \SI{360}{\kelvin}$). For the $C$-$V$ and TAS measurements, an AGILENT 4294A was used (measurement frequency between \SI{40}{\hertz} and \SI{1}{\mega\hertz}, AC probing voltage \SI{50}{\milli\volt}). For TAS, no additional DC bias was applied. 

\section{Results}

\begin{figure}
	\centering
	\includegraphics[width=\columnwidth]{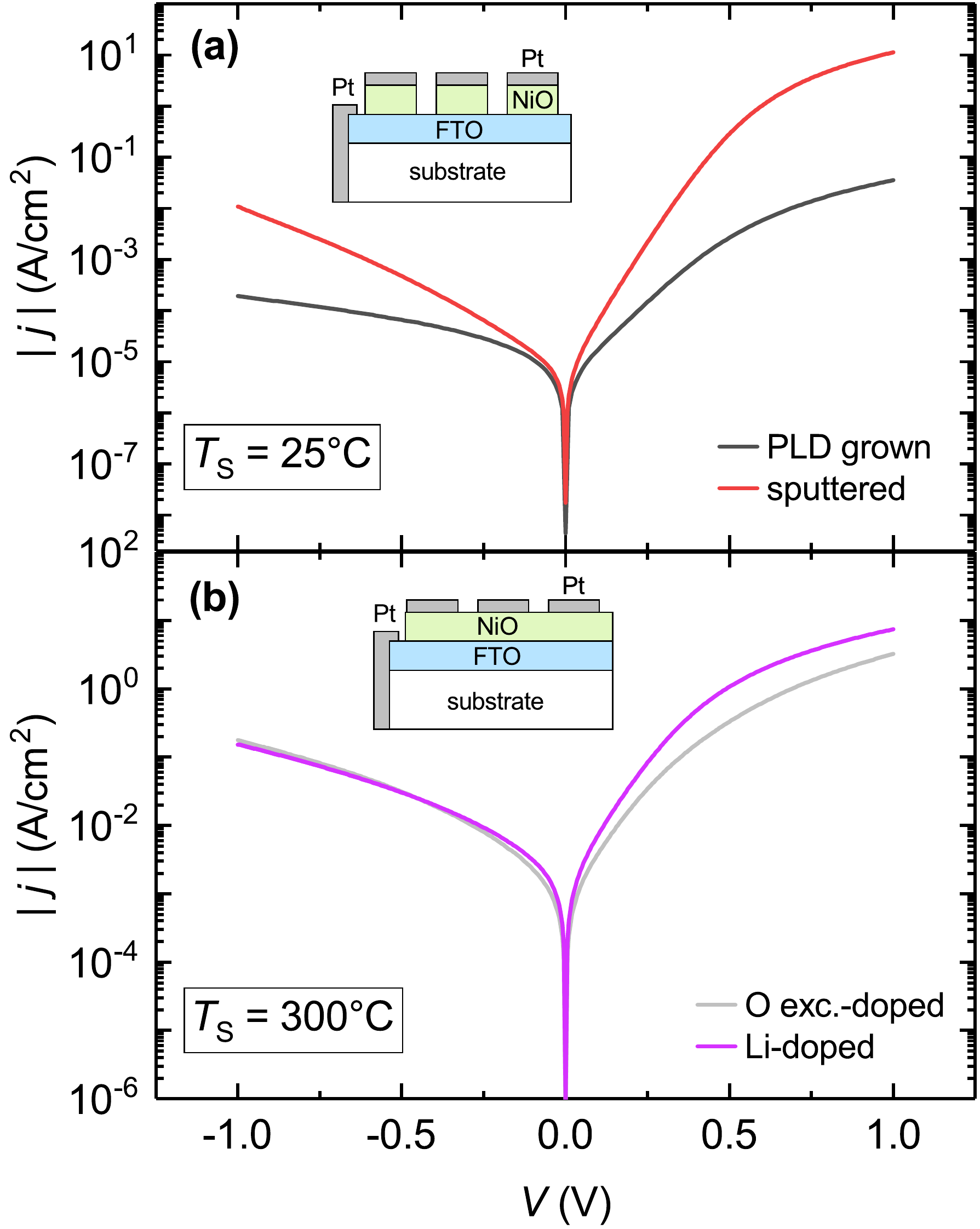} %
	\caption{Current density-voltage relations of different NiO/FTO structures. (a) NiO grown at room temperature, using PLD and magnetron sputtering, (b) grown at a substrate temperature of about \SI{300}{\degreeCelsius}, using oxygen excess or Li incorporation for doping. Insets show cross-sectional schematics of the sample structure.}
	\label{fig:jV_all_NiO}
\end{figure}

Fig.~\ref{fig:jV_all_NiO} shows typical characteristics measured on the NiO/FTO heterostructures. The rectification ratio ${S\mlwr{R} = \left|\nicefrac{j(\SI{1}{\volt})}{j(\SI{-1}{\volt})}\right|}$ takes on values between \num{3} orders of magnitude for the magnetron-sputtered NiO and \num{1.3} orders of magnitude for the high temperature-grown, oxygen excess-doped NiO. Based on the general diode equation

\begin{eqnarray}
 j = j\mlwr{s}\left[\exp\left(\frac{eV}{n\mlwr{id}k\mlwr{B}T}\right)-1\right],
\end{eqnarray}

with $j\mlwr{s}$ saturation current density and $n\mlwr{id}$ ideality factor of the diode, $e$ elementary charge, $k\mlwr{B}$ Boltzmann's constant and $T = \SI{300}{\kelvin}$ measurement temperature, diode ideality factors between \num{1.7} for magnetron sputttered NiO and \num{2.8} for room temperature PLD-grown NiO can be derived. In the case of high temperature-grown NiO, a monoexponential increase of the current density cannot be discerned at all. For an ideal, one-sided heterojunction where one junction partner is considerably higher doped than the other, which is shown to hold in this case further below, an ideality factor close to 1 is expected. The deviation from this value is likely explainable by the multicontact model developed by Johnson, Smith and Yearian \cite{Johnson.1950}. In this framework, diode areas with laterally varying properties, like local built-in potential and electrical conductivity, can give rise to such observations. We infer that lateral inhomogeneities of the NiO layer doping are responsible for the values of $n\mlwr{id}$.

\begin{figure}
	\centering
	\includegraphics[width=\columnwidth]{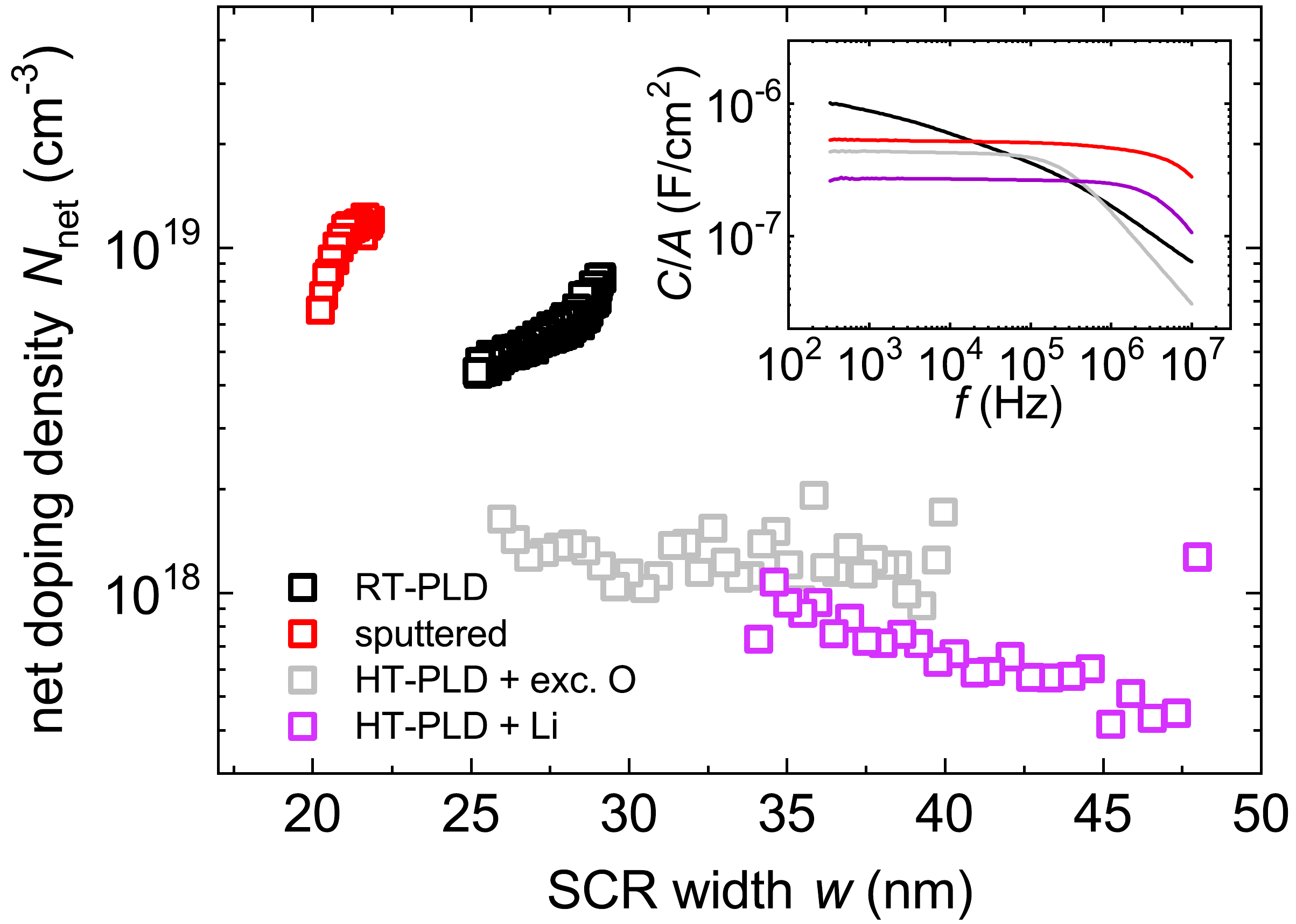}
	\caption{Charge concentration profiles of NiO/FTO heterostructures with differently deposited NiO layers, as obtained from $C$-$V$ measurements. Inset: frequency dependence of the capacitance per area at $V = \SI{0}{\volt}$.}
	\label{fig:doping_profiles_all_NiO}
\end{figure}

In Fig.~\ref{fig:doping_profiles_all_NiO} profiles of the net doping density $N\mlwr{net}$ with varying space charge region width $w$ from all four NiO film types are displayed, as obtained by means of capacitance-voltage ($C$-$V$) measurements, using the relations

\begin{align}
N\mlwr{net} &= -\frac{2}{e \varepsilon\mlwr{r}\varepsilon\mlwr{0} A^2} \left(\frac{\diff}{\diff V} C\mupr{-2}\right)\mupr{-1}, \\
w &= \frac{\varepsilon\mlwr{s}\varepsilon_0 A}{C}
\end{align}

where $\varepsilon\mlwr{r}\varepsilon_0 = 11.6\varepsilon_0$ is the NiO dielectric permittivity and $A$ the cross-sectional area of the contact. The voltage range used was between \SI{-0.5}{\volt} and \SI{0.1}{\volt}. The net doping ranges between some \SI{e17}{\per\cubic\centi\meter} for the Li-doped NiO, and slightly above \SI{e19}{\per\cubic\centi\meter} for the sputtered NiO. The inset of Fig.~\ref{fig:doping_profiles_all_NiO} shows the frequency dependence of the capacitance for all four sample types. It becomes apparent that for the sputtered and the Li- and O excess-doped HT-grown films, a broad spectral region exists where $C$ is constant down to the lowest fequency measurable with our setup (\SI{40}{\hertz}). Only for the film grown by PLD at room temperature, the frequency dependence exhibits a continuous decrease across the entire frequency range. Such a behavior is often characteristic of a broad distribution of electronic states participating in carrier displacement. Under AC electric fields at low frequencies, electronic charges are being periodically exchanged between states below and above the Fermi level. Increasing the frequency of the AC field leads to a progressive cut-off of higher-lying states, such that only states with energies closer to the Fermi level can participate. As a result, the dynamically determined capacitance is a decreasing function of the measurement frequency. For the Li- and O excess-doped as well as the sputtered NiO, the distribution of density of states appears to be rather narrow, as exemplified by the almost constant capacitance up to the \si{\mega\hertz} range. For the RT-PLD film, the dispersion is indicative of a broad distribution of electronic states, in accordance with the results from our earlier work \cite{Karsthof.2019}. The charge density profile for that sample, as shown in Fig.~\ref{fig:doping_profiles_all_NiO}, must therefore be considered  a lower boundary for the actual carrier concentration. For the other samples the charge concentrations can be considered more reliable.

Another important point to be noticed from the data in Fig.~\ref{fig:doping_profiles_all_NiO} is the fact that doping with O excess achieve higher carrier concentrations in room-temperature processes than for high-temperature growth. This is, again, in accordance with our earlier work \cite{Karsthof.2020} and also with the results from Gutiérrez \textit{et al.} \cite{Gutierrez.2020}: elevated temperatures cause the Ni vacancy acceptors brought into the film under non-equilibrium conditions to dissolve by the release of oxygen, thereby returning to a concentration that is closer to the thermodynamic equilibrium value. This process already occurs during growth and leads to NiO films grown at higher temperature having lower carrier concentration than the ones grown at room temperature, even though the same oxygen partial pressures were used. In the further course of this paper, we will focus on the high-temperature grown samples. This is, in part, due to higher stability of the defect population herein, but also because of a higher degree of disorder in the room-temperature fabricated films which makes a unique assignment of defect signatures difficult.

\begin{figure}
	\centering
	\includegraphics[width=\columnwidth]{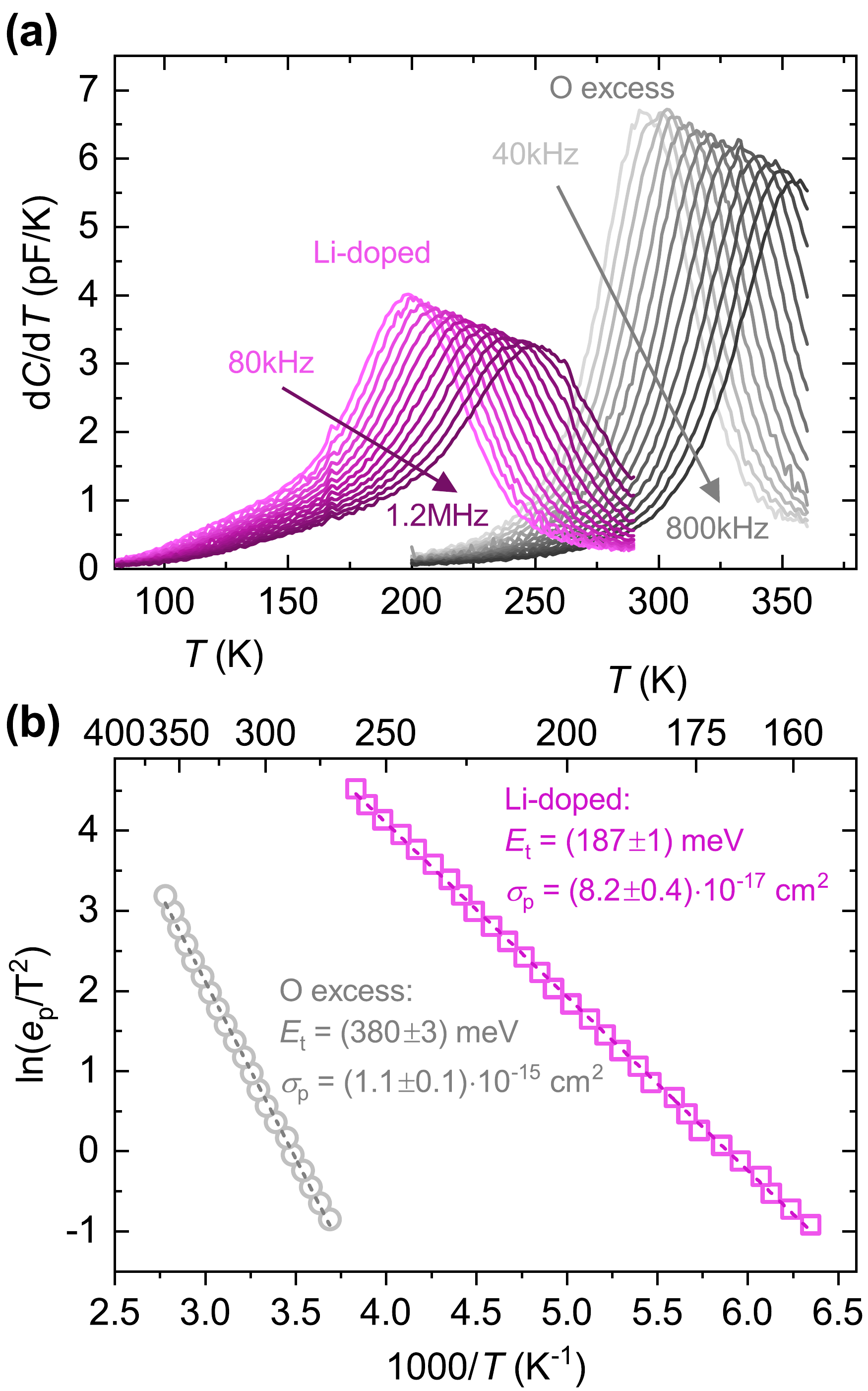}%
	\caption{Thermal admittance spectroscopy results for selected NiO/FTO diodes on oxygen-excess grown and Li-doped NiO films. (a) Differential capacitances, (b) extracted emission rates vs. temperature, including fits according to the Arrhenius relation (Eqn.~(\ref{eq:TAS_activation})).}
	\label{fig:TAS_dCdT_Arrhenius}
\end{figure}

Thermal admittance spectroscopy (TAS) measures the frequency and temperature dependence of the complex admittance $\underline{Y}$. A typical way of modelling admittance measurements of diodes is using an equivalent circuit consisting of a capacitor (capacitance $C$) and a parallel resistor (conductance $G$). In this case, $\underline{Y}= G + \imag \omega C$, with $\omega = 2\pi f$. The frequency dependence of both $G$ and $C$ is governed by the displacement of mobile charge carriers. For conventional band semiconductors, this consists of two contributions: free carriers at the edge of the space charge region are periodically moved back and forth within the (conduction or valence) band, while electronic defects with states within the semiconductor band gap can be filled and emptied during one period of the AC voltage as long as their energy level crosses the Fermi energy at some point within the SCR. Because charging and discharging of defects is tied to the emission and capture of mobile carriers from a nearby band, the rate of this process is limited. At a certain critical frequency of the applied AC voltage, carrier emission will be too slow to follow, and the charge state of the defect does not change anymore. This cutoff is visible in the spectra of both real and imaginary part of $\underline{Y}$: a step appears in $C$ (or equivalently, a peak in the derivative $\nicefrac{\diff C}{\diff T}$), while the quantity $\nicefrac{G}{\omega}$ exhibits a peak. The frequency $\omega\mlwr{0}$ at which the cutoff appears is related to the emission rate $e\mlwr{p}$ of the defect level, and can be described by (in the case of a $p$-type semiconductor)

\begin{align}
	\omega_0 \approx 2e\mlwr{p} &= 2\nu_0 \exp\left(-\frac{E\mlwr{t}}{k\mlwr{B}T}\right) = 2v\mlwr{th} \sigma\mlwr{p}N\mlwr{V}\exp\left(-\frac{E\mlwr{t}}{k\mlwr{B}T}\right) \nonumber \\
	&= 2\xi T^2 \exp\left(-\frac{E\mlwr{t}}{k\mlwr{B}T}\right), \label{eq:TAS_activation}
\end{align}

with $\nu_0$ being the escape attempt frequency, $E\mlwr{t}$ the distance of the trap level to the nearmost band, $v\mlwr{th}$ the thermal velocity of the free carriers, $\sigma\mlwr{p}$ the defect's capture cross section for holes and $N\mlwr{V}$ the effective density of states in the valence band \cite{Walter.1996}. Here, $v\mlwr{th} \propto T\mupr{1/2}$ and $N\mlwr{V} \propto T\mupr{3/2}$ are assumed. Equation (\ref{eq:TAS_activation}) shows that carrier emission is thermally activated with an energy $E\mlwr{t}$  the corresponding cutoff features in $C$ and $\nicefrac{G}{\omega}$ will also show a shift with measurement temperature. An Arrhenius plot of $\nicefrac{e\mlwr{p}}{T^2}$ then yields the depth $E\mlwr{t}$ of the trap as well as its capture cross section $\sigma\mlwr{p}$. In Fig.~\ref{fig:TAS_dCdT_Arrhenius}(a), the capacitance $C$ differentiated with respect to the temperature is shown for a selected diode on the Li-doped and O-excess NiO film as a function of temperature for a few measurement frequencies. A well-resolved peak appears for both samples, and apart from this one feature per sample type, no other peaks were observed within the accessible temperature range. Fig.~\ref{fig:TAS_dCdT_Arrhenius}(b) displays the temperature dependence of the associated emission rates of the peaks, calculated based on Equation~(\ref{eq:TAS_activation}), and a simple temperature activation becomes apparent. Three to four diodes were characterized per sample, and the data shown in Fig.~\ref{fig:TAS_dCdT_Arrhenius} are representative of typical results. By fitting the data based on Eqn.~(\ref{eq:TAS_activation}), and using values for the thermal hole velocity of the order of $v\mlwr{p,th} \approx \SI{e7}{\centi\meter\per\second}$ and $N\mlwr{V} \approx \SI{e19}{\per\cubic\centi\meter}$, trap energies of $E\mlwr{t,Li} = \SI{190\pm5}{\milli\electronvolt}$ and $E\mlwr{t,o-exc} = \SI{409\pm38}{\milli\electronvolt}$, and apparent hole capture cross sections of $\sigma\mlwr{p,Li} = \SI{6.5\pm1.6e-17}{\square\centi\meter}$ and $\sigma\mlwr{p,o-exc} = \SI{2.3\pm0.9e-15}{\square\centi\meter}$ could be extracted, respectively (values given are averages over all measured diodes, error includes statistical variation). It shall be noted at this point that the trap energies fall within the range of the ones given in Table~\ref{tab:lit_act_energies} for the $(0/-)$ charge transition levels of the major acceptors, Li\lwr{Ni} and V\lwr{Ni}.

\section{Discussion}

It is compelling to assign the features observed in TAS to ionization processes of the prominent acceptors, and as we will show in the following, this is likely correct. There are, however, two caveats to be discussed. First, being an AC measurement, TAS is capable of detecting processes that critically limit frequency of the charge transport processes to, from and around the defects. In conventional semiconductors possessing high free carrier mobilities, carrier capture and emission are the limiting processes in that regard. However, for a low-mobility material like NiO, this assumption should be critically reviewed. It was recently shown \cite{Karsthof.2019,Kokubun.2020} that the electronic transport in both Li-doped and Ni-deficient NiO can be modeled using a theoretical framework in which the charge carriers form polarons on the acceptor sites, and move through the crystal by hopping between closely spaced acceptors (polaronic interacceptor hopping, PIH). This process was shown to determine the DC conductivity of an undepleted, doped NiO film to a large extent. The question therefore is if the charge transfer observed by means of TAS in the present paper can be identified with the direct movement of carriers between acceptors, instead of excitation to the valence band and subsequent free carrier movement within that band. 

In such a case, the determined activation energy of the trap-like feature should be identical with that for DC transport in undepleted devices. To check this, we measured the activation energy of an undepleted, Li-doped NiO film, sandwiched between two Pt electrodes, similar to the method described in Ref.~\cite{Karsthof.2019}. $\sigma\mlwr{dc}$ turned out to follow an Arrhenius behavior with an activation energy of approximately \SI{370}{\milli\electronvolt} (See Fig.~S1 of the Supplemetary Material \cite{supplMat}). In the framework of the PIH transport model, this energy corresponds to the hopping barrier between two neighboring Li\lwr{Ni} acceptors. Using the data of Ref.~\cite{Kokubun.2020} as reference, an intersite separation larger than \SI{4}{\nano\meter}, or a Li\lwr{Ni} concentration of below \SI{0.03}{\atpercent}, can be inferred. This agrees well with the intended Li concentration and the net doping measured on our films. The process observed by TAS on an identically fabricated, but depleted, Li-doped NiO layer possessing an activation energy of \SI{190}{\milli\electronvolt}, as in Fig~\ref{fig:TAS_dCdT_Arrhenius}(b), can therefore not be related to the cut-off of the intersite hopping transport. For the sample grown under O excess, the acitivation energies for the TAS peak and the DC conductivity, the latter being around \SI{380}{\milli\electronvolt} as determined in Ref.~\cite{Karsthof.2019}, are very similar. However, a key ingredient of the PIH transport model is the critical dependece of the hopping barrier height on the intersite separation. Although being thermally considerably more stable than the room-temperature grown NiO films discussed in Ref.~\cite{Karsthof.2020}, even the O excess sample grown at elevated temperatures exhibited a slow decrease of V\lwr{Ni} acceptor density with time. After the sample had been stored at room temperature for about 18 months, renewed measurements revealed that the net doping of the film had decreased by a factor of 20, from \SI{1.4e18}{\per\cubic\centi\meter} to \SI{7e16}{\per\cubic\centi\meter} (inset in Fig.~S2 of the Supplementary Materials \cite{supplMat}). This is even lower than the net doping of the intentionally annealed, room-temperature grown samples studied in Ref.~\cite{Karsthof.2020}. It is therefore reasonable to assume that the intersite hopping barrier in the long-time room temperature-annealed film investigated in this work is at least as large as in the referenced samples, i.e. 700-\SI{800}{\milli\electronvolt}. However, renewed TAS measurements on the same, but aged, HT-grown sample, revealed that the rates associated with the TAS peak had not changed (Fig.~S2 in the Supplementary Materials). For that reason, it can be concluded that also in the O excess sample, the TAS feature is unrelated to a frequency cut-off of the PIH transport process.

To understand why interacceptor transport does not play any role for the AC capacitance signal of the depleted devices, it is insightful to consider what conditions would have to be met in order for an acceptor state to be ionized though intersite hopping alone. An increase in space charge only occurs when the total concentration of holes bound to the acceptors decreases, i.e. holes leave the NiO layer by inter-acceptor hopping until they reach the Pt back contact. This means that the carrier has to travel over a distance of the order of \SI{100}{\nano\meter} (from inside the space charge region to the NiO/Pt interface). Since this region is mainly field-free, the carrier movement would be diffusion-limited. Assuming an effective mobility for inter-acceptor hopping of the order of \SI{2e-7}{\square\centi\meter\per\volt\per\second} in line with Ref.~\cite{Karsthof.2019}, and based on the Einstein relation $D = \nicefrac{\mu k\mlwr{B}T}{e}$, the hole diffusivity at room temperature is roughly $D=\SI{5e-9}{\square\centi\meter\per\second}$. The hole transit through the entire quasi-neutral region of length $l\mlwr{qn} \approx \SI{100}{\nano\meter}$ therefore takes about $t\mlwr{diff} = \nicefrac{l\mlwr{qn}^2}{D} = \SI{20}{\milli\second}$ which is equivalent to a carrier removal rate through PIH of the order of \SI{50}{\per\second}, which is considerably lower than the measured emission rates. In other words, PIH does not offer hopping rates high enough to extract a significant amount of holes within one measurement cycle, compared to hole excitation from the trap to the valence band and extraction therein.

The next issue possibly arising when carriers are excited into the valence band is posed by low mobility within that band. In a recent publication, Wang \textit{et al.} explored the implications of low carrier mobility (i.e. hopping transport even in the extended states of the host crystal) on TAS spectra by means of numerical simulations \cite{Wang.2018}. The authors conclude that in such materials, a frequency cut-off originating from carrier mobility freeze-out, a so-called dielectric relaxation peak (DRP), may occur within the studied frequency range. Such a feature may be mistaken for a defect signature, and it may also superimpose other features that actually are related to a defect response if the hopping rate in the band is lower than the escape frequency from an acceptor. It has been speculated early on that the $d$ electronic character of valence band states in NiO leads to localization of holes, and transport was suggested to be due to a hopping mechanism even in band-like extended states \cite{Adler.1970}. Consequently, also the transport within the "valence band" may be via hopping. Therefore, the possibility of the occurence of dielectric relaxation in TAS measurements needs to be critically reviewed.

Wang \textit{et al.} found a DRP to dominate the TAS spectrum of a heterostructure comprising the organic semiconductor P3HT \cite{Wang.2018}. The work by Reisl\"{o}hner, Metzner and Ronning revealed that the freeze-out of the mobility of hopping carriers in Cu(In,Ga)Se\lwr{2} layers produces a trap-like feature \cite{Reisloehner.2010}. If the features observed by TAS in this work were caused by the freeze-out of hopping transport in the bands, the critical temperature would be independent of the used acceptor. This is in clear contrast to what is observed in Fig.~\ref{fig:TAS_dCdT_Arrhenius}(a): comparing the Li-doped and the O excess sample, the peak temperatures differ by about \SI{100}{\kelvin} for the same set of measurement frequencies. Such a strong dopant dependency is incompatible with the assumption of a cut-off of hopping transport in the bands within the frequency and temperature window used here. Therefore the observed features in TAS are not caused by a dielectric relaxation. 

\begin{figure}
	\centering
	\includegraphics[width=\columnwidth]{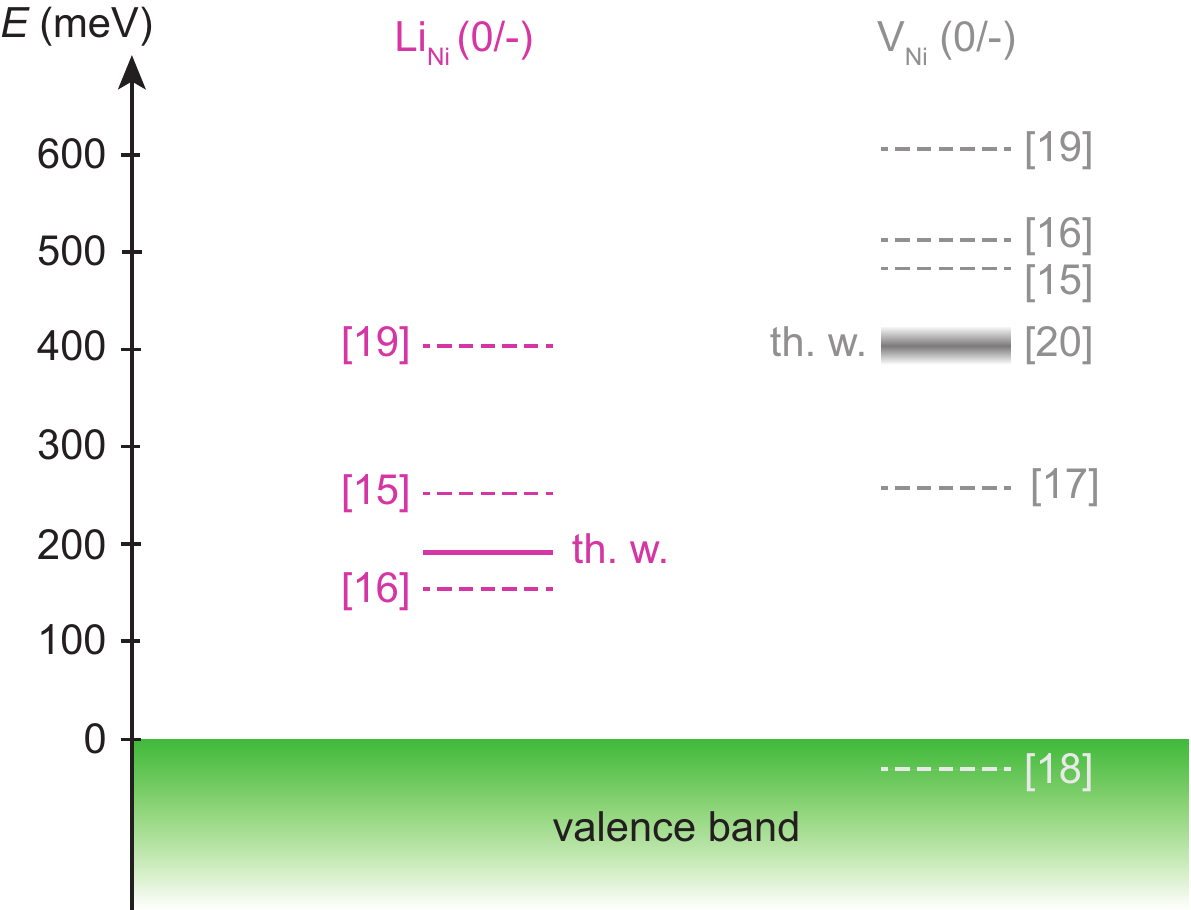}
	\caption{Energy level scheme of the Li\lwr{Ni} and V\lwr{Ni} $(0/-)$ charge transition levels as determined in this work, along with values from the literature.}
	\label{fig:trap_levels}
\end{figure}

Based on these points, we suggest the attribution of the traps detected by TAS to the $(0/-)$ charge transition levels of both the Li\lwr{Ni} and the V\lwr{Ni} acceptors. The energy level scheme is shown in Fig.~\ref{fig:trap_levels}, which also includes the literature values from Table~\ref{tab:lit_act_energies} for comparison. 

It is also interesting to note the large diffference in the values obtained for the apparent hole capture cross sections (\SI{6.5e-17}{\square\centi\meter} for the Li\lwr{Ni} and \SI{2.3e-15}{\square\centi\meter} for the V\lwr{Ni} acceptor, respectively). This may point towards fundamental differences in trap properties, like the spatial extension of the trap potential. As vacancies tend to locally deform the lattice to a larger extent than a substitutional ion with well-matched ion radius, like Li in NiO, one can expect the vacancy to create a higher amount of localized strain that could also extend the range of electrostatic forces in the vicinity of the trap.

A second ionization of the V\lwr{Ni}, the $(-/2-)$ CTL, has not been observed within our experiments. Since the value for the $(0/-)$ CTL from Ref.~\cite{Dawson.2015} matches our experimental one well, we use the value for the $(-/2-)$ level of \SI{600}{\milli\electronvolt} given in that reference to estimate the temperature window for an observation of the second ionization, assuming that the hole capture cross section is identical to that of the $(0/-)$ CTL. For the same set of frequencies shown in Fig.~\ref{fig:TAS_dCdT_Arrhenius}(a), the temperature window where the $(-/2-)$ CTL of the V\lwr{Ni} can be expected to be observed in TAS is \SI{435}{\kelvin} to \SI{525}{\kelvin}.

%

\section{Conclusion}

We have studied FTO/NiO $n\mupr{++}p$ heterostructures with Li-doped or O-rich NiO films, respectively. These structures are current-rectifying with a rectification ratio between one and two orders of magnitude at $\pm\SI{1}{\volt}$, indicating a depletion layer within the NiO. This structure enabled the characterization of trap levels by means of space charge spectroscopy. Using thermal admittance spectroscopy, we have found one dominant feature that can clearly be attributed to hole capture to and emission from one trap level per doping type. We tentatively attribute these levels to the $(0/-)$ charge transition levels of the two prominent acceptors Li\lwr{Ni} and V\lwr{Ni}, which are found at \SI{190\pm5}{\milli\electronvolt} and \SI{409\pm38}{\electronvolt} above the valence band edge, in compliance with literature values. The apparent hole capture cross sections of the traps have been determined to be \SI{6.5\pm1.6e-17}{\square\centi\meter} for the Li\lwr{Ni} and \SI{2.3\pm0.9e-15}{\square\centi\meter} for the V\lwr{Ni} acceptor, respectively.

\begin{acknowledgments}
	This work was funded by the Deutsche Forschungsgemeinschaft (DFG, German Research Foundation) -- project number 31047526, SFB762, project B06. Financial support was furthermore kindly provided by the Research Council of Norway and the University of Oslo through the frontier research project FUNDAMeNT (no. 251131, FriProToppForsk-program). The Research Council of Norway is acknowledged for the support to the Norwegian Micro- and Nano-Fabrication Facility, NorFab, project number 245963.
\end{acknowledgments}

\section*{Data Availability Statement}
The data that support the findings of this study are available from the corresponding author upon reasonable request.

\bibliography{ref}

\end{document}